\documentstyle[twoside,fleqn,espcrc2,epsf]{article}

\newcommand{\AmS}{{\protect\the\textfont2
  A\kern-.1667em\lower.5ex\hbox{M}\kern-.125emS}}

 \title{Correlation functions between topological objects -- \\ 
        field theoretic versus geometric definitions} 

\author{Markus 
Feurstein\addtocounter{address}{-1}\address{Institut f\"ur Kernphysik,
Technische Universit\"at Wien, A--1040 Vienna, Austria},
Harald Markum and Stefan Thurner}

\begin{document}

\begin{abstract}
We analyze topological objects in pure gluonic $SU(2)$ lattice gauge
theory and compute correlation functions between instantons and
monopoles. Concerning the instantons we use geometric and field theoretic
definitions of the topological charge. On a $12^3\times 4$ lattice 
it turns out that topological quantities have a non-trivial 
local correlation. The auto-correlation functions of the topological charge
depend on cooling for both definitions. We fit the  
correlation functions to exponentials and obtain screening masses. 
\end{abstract}
\maketitle

{\it 1. Introduction and Theory.}
Classical gauge field configurations with a non-trivial topology
are believed to play an essential role in the confinement mechanism.
In the scenario of the dual 
superconductor abelian monopoles condense leading to confinement.
Large and interacting instantons could also produce confinement if they 
form an instanton liquid. Because of the distinctness of 
these two 
pictures the interesting question arises, whether instantons and monopoles 
are related to each other.
In a recent study we computed correlation functions between monopoles and 
instantons using naive definitions of the topological charge \cite{lat95}.
We found a non-trivial local correlation. Now we investigate if this also
holds for a geometric charge definition. 

There exist several definitions of the  topological charge on the lattice.
We use a field theoretic and a geometric charge definition. The 
field theoretic prescription is a straightforward  discretization of
the continuum expression.
To get rid of the renormalization 
constants we apply the ``Cabbibo-Marinari cooling method'' with a cooling 
parameter $\delta = 0.05$. 
The geometric charge definitions interpolate the discrete set of link 
variables to the continuum and then calculate the topological charge 
directly. 
In our studies in $SU(2)$ we employ the hypercube prescription for the 
field theoretic definition \cite{divecchia} and 
 the locally gauge invariant L\"uscher charge 
definition \cite{schier_l}.

After fixing the gluon field to the maximum abelian gauge,
abelian color magnetic monopole 
currents $m(x,\mu)$ are evaluated over elementary cubes \cite{kronfeld}. 
The quantity of further interest is the monopole density being  
defined as $ \rho(x) = \frac{1}{ 4V_{4}} \sum_{\mu} | m(x,\mu) |$.

We calculate~the~cor\-relation func\-tions \linebreak[4]
$\langle q(0) q(r) \rangle$  and $\langle |q(0)|\rho(r) \rangle$   
for the geometric and the field theoretic definition of the topological 
charge and  normalize them  to one after subtracting 
the corresponding cluster values.

{\it 2. Results.}
Our simulations were performed on a $12^{3} \times 4$ lattice with
periodic boundary conditions using the Metropolis algorithm.
The observables were studied in pure $SU(2)$ both in the confinement
and the deconfinement phase at inverse gluon coupling
$\beta=4/g^{2}=2.25$ and $2.4$, respectively, employing the 
Wilson plaquette action. 
 For each run  we made 10000 iterations and measured our
observables
after every 100th iteration. Each of these 100 configurations was
first cooled and then
subjected to 300 gauge fixing steps enforcing the maximum abelian gauge.

The auto-correlation functions of the topological charge densities for the 
hypercube definition (l.h.s.) and for the L\"uscher
 method (r.h.s.) are displayed in 
Fig.~1 for 0, 5, 20 cooling steps in the confinement phase at $\beta=2.25$.
Without cooling both auto-correlation functions are $\delta$-peaked 
due to the dominance of  quantum fluctuations.
The auto-correlations become broader with cooling reflecting the existence 
of extended instantons, whose core sizes behave rather similar in the 
deconfinement phase at $\beta=2.4$. 
\clearpage 
%
\begin{figure*}[h]
\begin{center}
\begin{tabular}{cc}
\hspace{0.4cm} Field theoretic definition  &
\hspace{0.5cm} Geometric definition \vspace{0.5cm}\\
\epsfxsize=6.0cm\epsffile{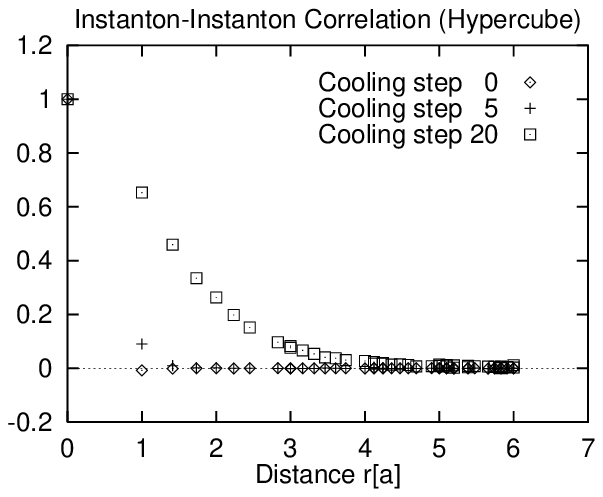}
&
\epsfxsize=6.0cm\epsffile{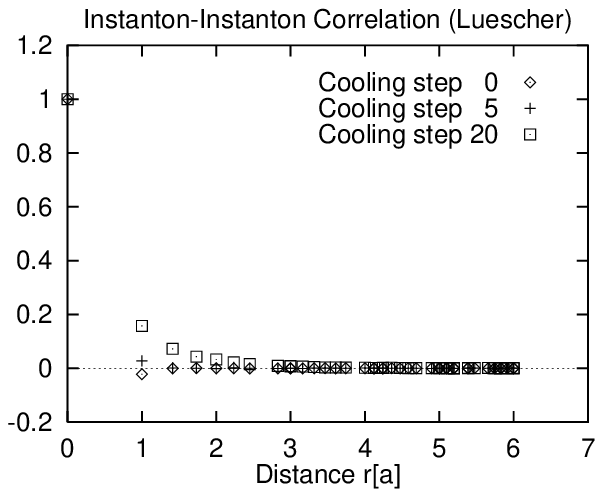} \\ \\
\end{tabular}
\end{center}
\vspace{-0.5cm}
{\baselineskip=13pt
Figure 1.~Auto-correlation functions of the topological charge density 
using the hypercube definition (l.h.s.) and the L\"uscher definition (r.h.s.)
in the confinement phase for 0, 5, 20 cooling steps.
Both auto-correlations grow with 
cooling indicating the existence of extended instantons.  
\baselineskip=13pt}
\\
\begin{center}
\begin{tabular}{cc}
\hspace{0.3cm}  Confinement & 
\hspace{0.3cm}  Deconfinement\vspace{0.5cm}\\
\epsfxsize=6.0cm\epsffile{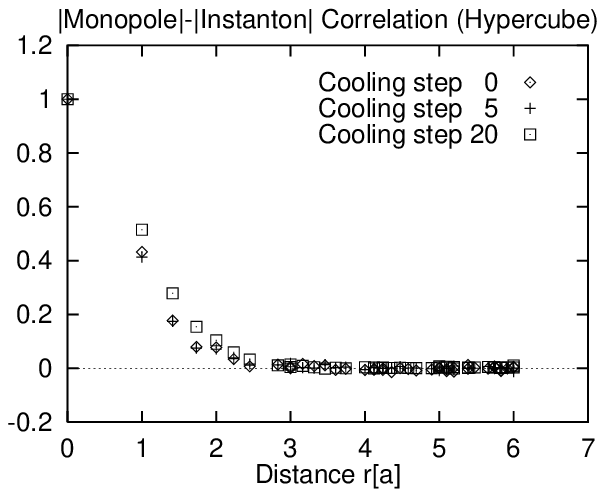}
&
\epsfxsize=6.0cm\epsffile{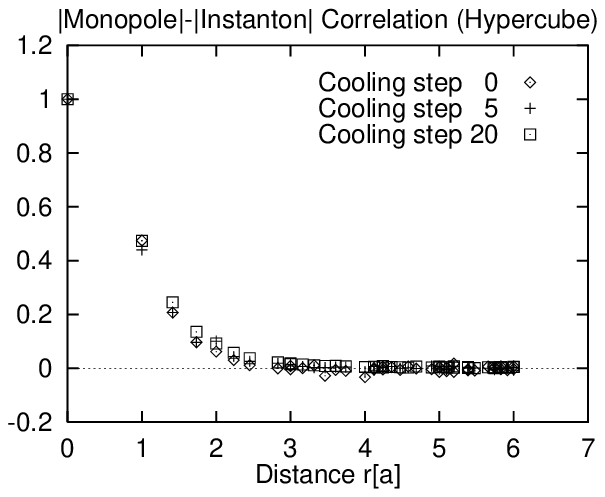} \\ \\
\epsfxsize=6.0cm\epsffile{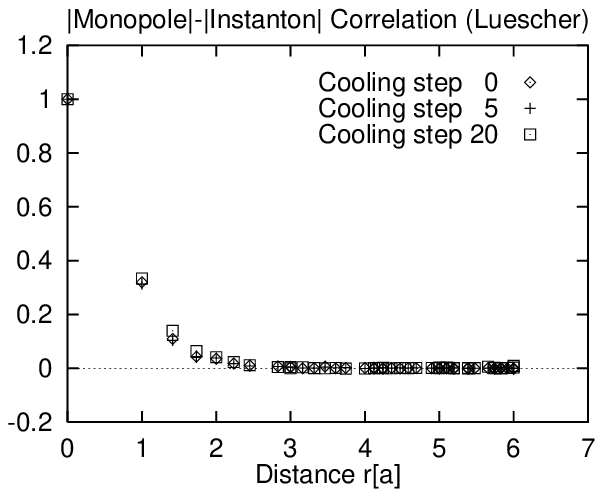}
&
\epsfxsize=6.0cm\epsffile{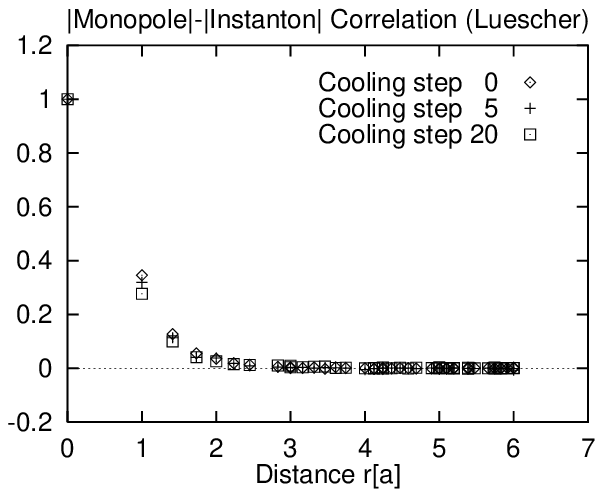} \\ \\
\end{tabular}
\end{center}
\vspace{-0.5cm}
{\baselineskip=13pt
Figure 2.~Correlation functions between the monopole density and the absolute 
value of the topological charge density for the hypercube definition and the 
L\"uscher definition in the confinement (l.h.s.) and the deconfinement phase 
(r.h.s.) for 0, 5, 20 cooling steps. For both definitions  the 
monopole-instanton  correlation functions are almost 
invariant under cooling and extend over approximately two lattice spacings.  
The correlation functions hardly change across the phase transition. 
\baselineskip=13pt}
\end{figure*}
\clearpage 

In Fig.~2 the correlation functions between  the monopole density and the
absolute value  
of the topological charge density for the hypercube and the L\"uscher 
definition are depicted in the confinement (l.h.s.) and the deconfinement 
phase (r.h.s.) for several cooling steps. Both definitions yield qualitatively 
the same result.  
For each charge definition
the correlation functions  are almost independent of 
cooling and extend over approximately two lattice units. 
This indicates that 
there exists a non-trivial local correlation between these topological objects
and that the probability for finding monopoles around instantons is clearly 
enhanced. The $\rho |q|$-correlations seem to be hardly influenced by the phase 
transition.  
\begin{table}[h]
\vspace{-1.2cm}
\begin{center}
\begin{tabular}{c}
 Confinement  ($\beta= 2.25$) \vspace{0.1cm}  \\
\begin{tabular}{lccc}
\hline
 Correlation &  Cool step 5  & Cool step 20  \\
\hline
$q-q$ (L\"u)          & ($1.16 \pm 0.59$) & $1.52 \pm 0.40 $   \\
$q-q$ (hyp)         & ($1.51 \pm 1.12$) & $1.10 \pm 0.48 $   \\
$|q|-\rho$ (L\"u)   & $2.03 \pm 0.22$ & $2.46 \pm 0.19 $   \\
$|q|-\rho$ (hyp)  & $1.86 \pm 0.15$ & $1.94 \pm 0.26 $   \\
\hline
\end{tabular} \vspace{0.3cm}\\ 
 Deconfinement  ($\beta= 2.4$) \vspace{0.1cm}  \\
\begin{tabular}{lccc}
\hline
 Correlation & Cool step 5 & Cool step 20 \\
\hline
$q-q$ (L\"u)          & ($1.94 \pm 0.71$) & $1.00 \pm 0.29 $   \\
$q-q$ (hyp)         & ($2.00 \pm 0.80$) & $1.09 \pm 0.31 $   \\
$|q|-\rho$ (L\"u)   & $1.69 \pm 0.51$ & $1.22 \pm 0.17 $   \\
$|q|-\rho$ (hyp)  & $1.86 \pm 0.32$ & $1.56 \pm 0.20 $   \\
\hline
\end{tabular}
\end{tabular} 
\end{center}
{\baselineskip=13pt
Table 1.~Screening masses from fits to 
exponential decays of the correlation functions between topological 
charge densities and between monopoles and instantons in the confinement and 
deconfinement phase for 5 and 20 cooling steps.
The numbers in brackets are not
reliable due to  bad S/N ratio.
\baselineskip=13pt}
\vspace{-0.7cm}
\end{table}

To gain a more quantitative insight, 
we analyze the correlation functions discussed above by fitting them 
to an exponential function. 
The resultant screening masses in lattice units 
are presented in Table~1 
in both phases for 5 and 20 cooling steps. 
The screening masses computed from the 
correlations between monopoles and instantons turn out  not very  
sensitive to cooling and to the phase transition.
 The error bars of the masses are large reflecting the large errors 
in the raw data not shown for clarity of plots.

Finally we add in Table 2 a result concerning  the ratio $R$ 
of spatial to time-like monopole densities as a function of cooling.
It has been reported  that this quantity sharply decreases across the
 deconfinement phase transition and that it migth serve as a reasonable
order parameter \cite{asym}. We observe that the same quantity also
decreases as a function of cooling yielding some      
doubt on the quality of this quantity as an order parameter.
\vspace{-1cm}
\begin{table}[h]
\label{moncool}
\begin{center}
\begin{tabular}{ c  c  c}
\hline
 Cool step  
             & $\beta$ = 2.25 &  $\beta$ = 2.40          \\
\hline
0  &  0.996 & 0.898  \\
5  &  0.975 & 0.571  \\
20 &  0.268 & 0.037 \\
\hline
\end{tabular}
\end{center}
{ Table~2. Ratio of spatial to time-like monopole density
  $R= \frac{\rho_s}{3 \rho_t}$
in the confinement and deconfinement for various cooling steps.}
\end{table}

{\it 3. Conclusion.}
We calculated correlation functions between monopoles and instantons and 
found a non-trivial local correlation indicating an enhanced probability
for finding monopoles around instantons. This finding does not depend on
the topological charge definition used and on the phase of the theory.  
 For a deeper understanding of this relationship we visualize 
monopoles and instantons for specific gauge field configurations 
(see these proceedings).

{\it 4. Acknowledgments.}
We thank Gerit Schierholz who provided us with a routine to compute 
the L\"uscher charge.
This work was partially supported by FWF 
      under Contract No.~P11456-PHY. The paper was presented by M.~F. 

\end{document}